\documentclass[12pt]{article}
\usepackage{amsfonts}
\usepackage{amssymb}
\usepackage{amsbsy}
\usepackage{graphics}

\input epsf.sty

\newlength{\TZ}
\newcommand{\BEQ}{\begin{equation}}     
\newcommand{\BEA}{\begin{eqnarray}}
\newcommand{\EEQ}{\end{equation}}       
\newcommand{\EEA}{\end{eqnarray}}
\newcommand{\D}{{\rm d}}                
\newcommand{\II}{{\rm i}}               
\newcommand{\demi}{\frac{1}{2}}         
\newcommand{\wht}[1]{\widehat{#1}}      
\newcommand{\lap}[1]{\overline{#1}}     

\renewcommand{\vec}[1]{\boldsymbol{#1}} 

\newcommand{\appsektion}[1]{\setcounter{equation}{0} \section*{Appendix #1}
\renewcommand{\theequation}{A\arabic{equation}}
              \renewcommand{\thesection}{A} }

\catcode`\@=11
\def\numberbysection{\@addtoreset{equation}{section}
        \def\theequation{\thesection.\arabic{equation}}}

\def\up#1{\raise 1ex\hbox{\sevenrm#1}}
\def\build#1_#2^#3{\mathrel{\mathop{\kern 0pt#1}\limits_{#2}^{#3}}}

\numberbysection

\begin{document}

\begin{titlepage}


\vskip 1.5 cm
\begin{center}
{\Large \bf Absence of logarithmic scaling in the ageing behaviour \\ 
of the $4D$ spherical model}
\end{center}

\vskip 2.0 cm
\centerline{ {\bf Maximilian Ebbinghaus},$^a$\footnote{Adresse after 
the 1$^{\rm st}$ of July 2007: Theoretische Physik, Universit\"at des Saarlandes, 
Postfach 151150, Geb\"aude E2.6, D -- 66041 Saarbr\"ucken, Germany} 
{\bf H\'el\`ene Grandclaude}$^a$ and {\bf Malte Henkel}$^{a,b}$ }
\vskip 0.5 cm
\centerline {$^a$Laboratoire de Physique des 
Mat\'eriaux,\footnote{Laboratoire associ\'e au CNRS UMR 7556} 
Universit\'e Henri Poincar\'e Nancy I,} 
\centerline{ B.P. 239, 
F -- 54506 Vand{\oe}uvre l\`es Nancy Cedex, France}
\centerline{$^b$ Centro de F\'{\i}sica T\'eorica e Computacional, 
Complexo Interdisciplinar da Universidade de Lisboa,}
\centerline{Av. Prof. Gama Pinto 2, P -- 1649-003 Lisboa Codex, Portugal}

\begin{abstract}
The non-equilibrium dynamics of the kinetic spherical model with a
non-conserved order-parameter, quenched to $T\leq T_c$ from a fully disordered 
initial state, is studied at its upper critical dimension
$d=d^*=4$. In the scaling limit where both the waiting time $s$ and the observation 
time $t$ are large and the ratio $y=t/s>1$ is fixed, 
the scaling functions of the two-time autocorrelation and
autoresponse functions do not contain any logarithmic correction factors
and the typical size of correlated domains scales for large times
as $L(t)\sim t^{1/2}$. 
\end{abstract}

\end{titlepage}

\section{Introduction}

The study of collective non-equilibrium behaviour has received a lot of
interest in recent years \cite{focus}. 
One particular aspect is the study of {\em ageing} 
which for example may arise if a many-body system is rapidly brought out of some
initial equilibrium state by the change of external control parameters such
as the temperature or an external field. If the change of these variables 
is such that the equilibrium state of the system is either at an equilibrium
critical point or else in a coexistence phase with at least two equivalent
but distinct equilibrium states, then one may observe (i) a slow, 
non-exponential relaxation, (ii) a breaking of time-translation-invariance
and (iii) dynamical scaling which are the three defining properties of 
ageing systems, see \cite{Bray94,Godreche02,Calabrese05,Henkel07} for 
reviews.\footnote{Since each of these three properties can be found
alone in situations where one would not speak of ageing, it seems reasonable
to insist on their simultaneous presence in ageing systems. For example, in
any system away from its stationary state time-translation-invariance will
not hold.} While ageing was first studied in glassy systems \cite{Struik78}, 
it has been 
realised more recently that its essential properties can also be found in 
simple ferromagnets, which may be easier to analyse and still may provide
useful clues for the understanding of the more complex glassy systems. 
In particular, in ferromagnets the form of dynamical scaling is quite simple
since the linear domain size scales algebraically according to 
$L(t)\sim t^{1/z}$. Ageing is conveniently
studied through the behaviour of two-time observables. For the
two-time autocorrelation and (linear) autoresponse functions, the most simple
kind of dynamical scaling behaviour is 
\BEA
C(t,s) &=& \langle \phi(t,\vec{r})  \phi(s,\vec{r}) \rangle
\:=\: s^{-b} f_C(t/s) 
\nonumber \\
R(t,s) &=& \left.\frac{\delta\langle\phi(t,\vec{r})\rangle}{\delta h(s,\vec{r})}
\right|_{h=0} \:=\: s^{-1-a} f_R(t/s)
\label{cr}
\EEA
where $\phi(t,\vec{r})$ is the order-parameter at time $t$ and at the location
$\vec{r}$, $h$ is the conjugate magnetic field, $a$ and $b$ are ageing 
exponents herewith defined and $f_{C,R}(y)$ are scaling functions. These scaling
forms are expected to hold when both times $t,s$ are large and their 
ratio $y=t/s>1$ is fixed. 

The scaling (\ref{cr}) is generally valid for systems with a so-called
simple ageing behaviour, but there are exceptions. In certain cases, the above
scaling forms are modified by additional logarithmic factors which already
manifests themselves in the scaling of the linear domain size, i.e. 
$L(t) \sim (t/\ln t)^{1/2}$ for a non-conserved order-parameter, 
and occurs in many systems where topological defects (e.g. vortices) play a role, 
such as planar magnets, frustrated spin systems, 
liquid crystals or superconductor arrays, see
\cite{Bray95,Puri95,Rojas99,Bray00,Berthier01,Jeon03,Abriet04,Dutta05,Schehr05,Schehr06,Lei07,Walter07}. 
Recently, for {\em equilibrium} critical
phenomena, Kenna, Johnston and Janke \cite{Kenna06a,Kenna06b} have constructed
a systematic theory for these logarithmic factors, 
based on an analysis of the complex zeroes of the
partition function, by which they derive scaling relations between the exponents
describing eventual logarithmic correction factors to simple scaling. They
checked the results of their analysis by comparing 
with the many results available for
logarithmic contributions in equilibrium scaling (see \cite{Berche07} for a 
very recent example). On the other hand, for non-equilibrium dynamical scaling
much less is known on possible logarithmic factors which makes tests of more
general ideas difficult. For that reason, we consider in this work the
ageing and dynamical scaling of the four-dimensional spherical model whose
logarithmic corrections factors to equilibrium scaling are well-established 
textbook knowledge \cite{Yeomans92}. While the scaling of its two-time
functions has been analysed in great detail,
starting with \cite{Godreche00b}, for either $d<4$ or $d>4$, to the best of 
our knowledge, the case $d=4$ has never been explicitly studied. 

In the next section, we define the model and briefly recall those elements
of the solution which we need for our analysis which is presented in section 3.
We conclude in section 4. Technical details of the calculation are given in
an appendix.

\section{Model and formalism}

The spherical model may be defined in terms of real spin variables
$S_{\vec{r}}\in\mathbb{R}$ attached to the sites $\vec{r}$ of a hyper-cubic 
lattice 
$\Lambda\subset \mathbb{Z}^d$ and which obey the mean spherical constraint
\BEQ
\sum_{\vec{r}\in\Lambda} S(t,\vec{r})^2 = {\cal N}
\EEQ
where $\cal N$ is the total number of sites of the lattice. 
The dynamics is assumed to be given by the stochastic Langevin equation
\BEQ
\partial_t S(t,\vec{r}) = \Delta S(t,\vec{r}) +\mathfrak{z}(t) S(t,\vec{r})+
\eta(t,\vec{r})
\EEQ
where $\Delta$ is the Laplacian with respect to $\vec{r}$, 
$\mathfrak{z}(t)$ is a Lagrange multiplier chosen such that the
spherical constraint holds and $\eta(t,\vec{r})$ is a centred gaussian noise
with variance 
$\langle\eta(t,\vec{r})\eta(t',\vec{r}')\rangle
=2T \delta(t-t')\delta(\vec{r}-\vec{r}')$. 
In writing this, a choice of units was made such that the corresponding kinetic
coefficient is set to unity. Throughout, we shall assume a fully disordered
initial state (where in particular the order-parameter vanishes) 
described by a gaussian variable and the variance 
$\langle S(0,\vec{r})S(0,\vec{r}')\rangle = \delta(\vec{r}-\vec{r}')$.\footnote{See
\cite{Fusco03} for a careful analysis of the mean spherical constraint. 
In the case of a non-vanishing initial magnetisation, the fluctuations in the Lagrange 
multiplier $\mathfrak{z}(t)$ must be taken into account \cite{Annibale06}. This
makes the solution of the model, even for $d\ne 4$, a formidable task. In the 
interest of a relatively simple presentation, we concentrate on the case of
a vanishing initial magnetisation {\em only}, where the formalism at hand with 
a non-fluctuating $\mathfrak{z}(t)$ is sufficient \cite{Annibale06}. See 
\cite{Annibale06,Calabrese06} for a detailed discussion of the ageing behaviour
when, starting from an ordered state, the quench is made to $T=T_c$.} 

Since the solution of this model is by now standard, 
see e.g. \cite{Godreche00b}, 
we shall merely quote those results which we shall need for our analysis
of the $4D$ case. One of the central quantities needed for the
analysis is the function $g(t) := \exp(2\int_0^t \!\D u\, \mathfrak{z}(u))$,
which, as a consequence of the spherical constraint, satisfies the Volterra
integral equation
\BEQ \label{V}
g(t) = f(t) + 2T \int_0^t \!\D t'\: f(t-t') g(t')
\EEQ
where the auxiliary function $f(t)$ is given by
\BEQ
f(t) = \int_{\cal B} \frac{\D \vec{q}}{(2\pi)^d}\: e^{-2\omega(\vec{q}) t} =
\left( e^{-4t} I_0(4t) \right)^d \;\stackrel{t\to\infty}{\sim} (8\pi t)^{-d/2}
\EEQ
where $\cal B$ denotes the first Brillouin zone, 
$I_0(u)=\frac{1}{\pi}\int_0^{\pi}\!\D\theta\, e^{-u\cos\theta}$ is a modified 
Bessel function \cite{Abra65} and the
dispersion relation is, for a hyper-cubic lattice, 
$\omega(\vec{q}) =\sum_{j=1}^d (2 - 2\cos q_j)$. In particular, the
critical temperature $T_c(d)>0$ for $d>2$ and is given by
\BEQ
\frac{1}{2 T_c(d)} = \int_0^{\infty} \!\D t\: \left( e^{-4t} I_0(4t) \right)^d
\EEQ
from which its numerical value is easily found for any $d$. For example,
$T_c(4)=6.45438\ldots$. Given the function $g(t)$,
any observable of interest is readily calculated. In what follows, we shall need
the single-time correlation function $\wht{C}_{\vec{q}}(t)$ given by
\BEQ
\langle \wht{S}_{\vec{q}}(t) \wht{S}_{\vec{q}'}(t)\rangle = 
(2\pi)^d \delta(\vec{q}+\vec{q}')\wht{C}_{\vec{q}}(t)
\EEQ
where the (discrete) Fourier transforms are defined by
\BEQ
\wht{f}_{\vec{q}} = \sum_{\vec{r}\in\Lambda} f(\vec{r}) 
e^{-\II \vec{q}\cdot\vec{r}} \;\; ; \;\; 
f(\vec{r}) = \int_{\cal B} \frac{\D \vec{q}}{(2\pi)^d}\: \wht{f}_{\vec{q}}
\: e^{\II \vec{q}\cdot\vec{r}}
\EEQ
One then has
\BEQ
\wht{C}_{\vec{q}}(t) = \frac{e^{-2\omega(\vec{q})t}}{g(t)} 
\left( 1+ 2T \int_0^t \!\D t'\: e^{2\omega(\vec{q})t'} g(t') \right)
\EEQ
where the initial condition $\wht{C}_{\vec{q}}(0)=1$ was used. 
In a similar fashion, one finds for the autocorrelation and autoresponse 
functions \cite{Godreche00b}
\BEA
C(t,s) &=& \left\langle S(t,\vec{r})  S(s,\vec{r}) \right\rangle \:=\: 
\frac{1}{\sqrt{ g(t) g(s)\,}} \left( f\left(\frac{t+s}{2}\right)
+ 2T \int_0^s \!\D t'\: f\left(\frac{t+s}{2}-t'\right) g(t')
\right) 
\nonumber \\
R(t,s) &=& 
\left.\frac{\delta\langle S(t,\vec{r})\rangle}{\delta h(s,\vec{r})}
\right|_{h=0} \:=\: 
f\left(\frac{t-s}{2}\right) \sqrt{\frac{g(s)}{g(t)}\,}
\label{CetR}
\EEA
Finally, we shall also measure the relevant length scale $L(t)$ of dynamical
scaling, which is also a measure of the linear size of correlated clusters,
by considering the normalised second moment of the single-time correlator
\BEA
L^2(t) &=& \frac{\sum_{\vec{r}\in\Lambda} \vec{r}^2 C(t,\vec{r})}
{\sum_{\vec{r}\in\Lambda} C(t,\vec{r})}  
\:=\: - \left. \frac{\partial^2 \wht{C}_{\vec{q}}(t) 
/\partial \vec{q}^2}{\wht{C}_{\vec{q}}(t)}\right|_{\vec{q}=\vec{0}} 
\nonumber \\
&=& 4d \:\frac{t+2T\int_0^t \!\D t'\:(t-t') g(t')}{1+2T\int_0^t \!\D t'\:g(t')}
\label{L2}
\EEA 

We remark that the response function, and also the correlators $C$ and the
typical length scale $L$ {\em at} criticality, where the thermal term dominates,
merely depend on ratios $g(t)/g(t')$ such that the global normalisation of
the function $g(t)$ will disappear from these physical observables. 

\section{Results}

\subsection{Solution of the Volterra equation}

In order to obtain $g(t)$ explicitly, one must solve the Volterra integral
equation (\ref{V}). Since its right-hand side has a convolution structure,
one considers the Laplace transformation 
$\lap{g}(p) = {\cal L}(g)(p)= \int_0^{\infty} \!\D t\, e^{-pt} g(t)$ and finds
\BEQ \label{glap}
\lap{g}(p) = \frac{\lap{f}(p)}{1-2T\lap{f}(p)}
\EEQ
{}From standard Tauberian theorems \cite{Montroll65}, the long-time behaviour 
of $g(t)$ follows from the asymptotics of $\lap{g}(p)$ as $p\to 0$. This in turn
requires the leading behaviour of $\lap{f}(p)$ which may be obtained in a 
well-known fashion, see \cite{Montroll65,Henkel84,Luck85}, which we outline
in the appendix. Together with the known results for $2<d<4$ and $d>4$ 
\cite{Godreche00b}, we have
\BEQ \label{f}
\lap{f}(p) = \frac{1}{2 T_c(d)} + \left\{
\begin{array}{ll} a_1 p^{d/2-1} & \mbox{\rm ~~;~ if $2<d<4$} \\
a_0 p( C_E + \ln p) & \mbox{\rm ~~;~ if $d=4$} \\
-a_2 p - a_1 p^{d/2-1} +\ldots & \mbox{\rm ~~;~ if $d>4$}
\end{array} \right.
\EEQ
where $a_0=(8\pi)^{-2}$, 
$a_1=-(8\pi)^{-d/2}|\Gamma(1-d/2)|$ for $d\ne 4$, 
$a_2=\int_{\cal B}{\frac{\D^d \mathbf{q}}{(2\pi)^d} 
\frac{1}{(2\omega(\mathbf{q}))^2}}=\int_0^{\infty} \!\D u\, 
u (e^{-2u} I_0(2u))^d$ and $C_E=0.5772\ldots$ is Euler's constant. 

\begin{figure}[h]
\centerline{
\epsfxsize=5.0in \epsffile{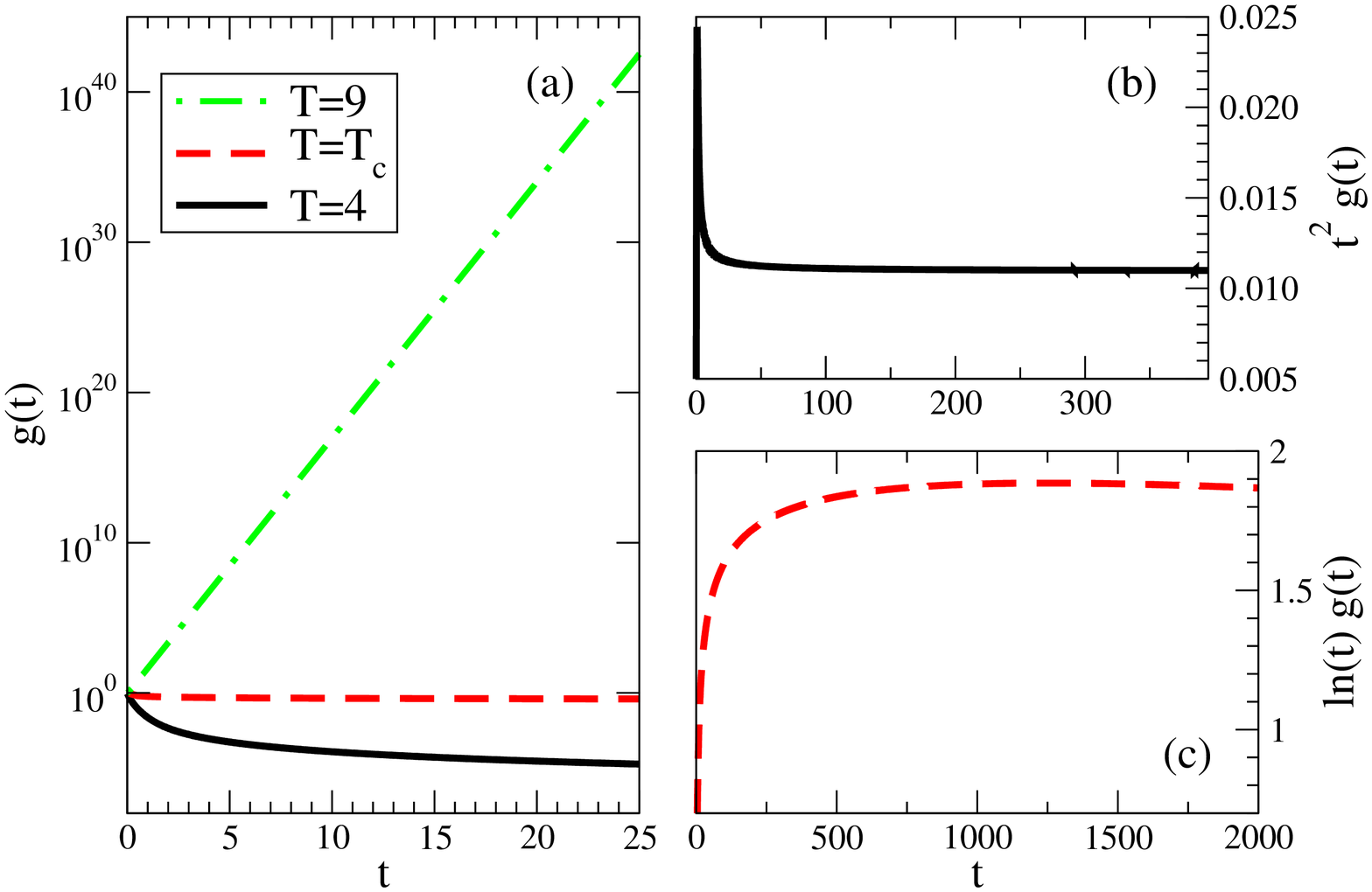}
}
\caption{(a) Behaviour of the function $g(t)$ in $d=4$ dimensions as a function 
of time, above, at and below $T_c(4)=6.45438\ldots$. In (b) the
behaviour $g(t)\sim t^{-2}$ for $T=4<T_c(4)$ and in (c) the behaviour 
$g(t)\sim \ln(t)^{-1}$ at $T=T_c(4)$ is illustrated.
\label{Abb1}
}
\end{figure}

\subsection{Scaling of autocorrelation and autoresponse}

We are now ready to discuss the long-time behaviour of our model where the three
cases $T>T_c$, $T<T_c$ and $T=T_c$ have to be distinguished. 
In figure~\ref{Abb1}a we illustrate the long-time
behaviour of $g(t)$ for these three cases, obtained by numerically solving
eq.~(\ref{V}) \cite{HeleMax}, for which $g(t)$ clearly shows
different behaviour. \\

\noindent \underline{\bf 1. $T>T_c$.} For quenches into the disordered phase, 
$g(t)$ is
obtained straightforwardly by the method of residues
\BEQ
g(t) = \frac{\lap{f}(1/\tau_{\rm eq}) e^{t/\tau_{\rm eq}}}
{-2T\lap{f}'(1/\tau_{\rm eq})} = \frac{16\pi^2}{T^2} 
\frac{e^{t/\tau_{\rm eq}}}{\ln(\tau_{\rm eq})-C_E-1}
\EEQ
where the finite relaxation rate is obtained from 
$\lap{f}(1/\tau_{\rm eq})=1/2T$ and reads for $T\gtrsim T_c$
$\tau_{\rm eq}\simeq\frac{1}{32\pi^2}\frac{T_c^2}{T-T_c}\ln\frac{T_c}{T-T_c}$.
Since the thermal equilibrium correlation length of the $4D$ spherical model
behaves as \cite{Kenna06a,Yeomans92} 
$\xi_{\rm eq}^2(T) \sim (T-T_c)^{-1}\ln (T-T_c)$ and 
one expects that the relaxation time $\tau_{\rm eq}\sim \xi_{\rm eq}^2$, 
this is consistent with known results. The two-time functions
relax within the finite time $\tau_{\rm eq}$ to their 
time-translation-invariant equilibrium values. \\

\noindent \underline{\bf 2. $T<T_c$}. 
Now, one has by combining (\ref{glap}) and (\ref{f}), for $p$ small enough  
\BEQ \label{gp}
\lap{g}(p) \simeq \left\{
\begin{array}{ll}
\frac{\displaystyle 1}{\displaystyle 2T_c M_{\rm eq}^2}
+\frac{\displaystyle a_1}{\displaystyle M_{\rm eq}^4} p^{d/2-1} &
\mbox{\rm ~~;~ if $2<d<4$} \\[1.6mm]
\frac{\displaystyle 1}{\displaystyle 2 T_c M_{\rm eq}^2}
+\frac{\displaystyle C_E}{\displaystyle 64\pi^2 M_{\rm eq}^4} p +
\frac{\displaystyle p\ln p}{\displaystyle
64\pi^2 M_{\rm eq}^4} & \mbox{\rm ~~;~ if $d=4$} \\[1.4mm]
\frac{\displaystyle 1}{\displaystyle 2 T_c M_{\rm eq}^2}
-\frac{\displaystyle a_2}{\displaystyle M_{\rm eq}^4} p
-\frac{\displaystyle a_1}{\displaystyle
M_{\rm eq}^4} p^{d/2-1} & \mbox{\rm ~~;~ if $d>4$}
\end{array} \right.
\EEQ
where the exact equilibrium result $M_{\rm eq}^2=1-T/T_c$ \cite{Yeomans92} 
was used. The presence of terms constant in $p$ in (\ref{gp}) signal that it is
not enough to concentrate on the asymptotic long-time behaviour of $g(t)$ but there are also `global' contributions to $g(t)$ which must be taken into account. 

Formally inverting the Laplace transform in (\ref{gp}) then leads 
to the following form for $g(t)$, 
\BEQ \label{gt}
g(t) \simeq \frac{\displaystyle 1}{\displaystyle 2T_c M_{\rm eq}^2} \delta(t) + \left\{
\begin{array}{ll}
\frac{\displaystyle f(t)}{\displaystyle M_{\rm eq}^4} &
\mbox{\rm ~~;~ if $2<d<4$} \\[1.6mm]
\frac{\displaystyle C_E}{\displaystyle 64\pi^2 M_{\rm eq}^4} \delta'(t) 
+ \frac{\displaystyle f(t)}{\displaystyle M_{\rm eq}^4} &
\mbox{\rm ~~;~ if $d=4$} \\[1.4mm]
(-1)\frac{\displaystyle a_2}{\displaystyle M_{\rm eq}^4} \delta'(t) 
+ \frac{\displaystyle f(t)}{\displaystyle M_{\rm eq}^4} &
\mbox{\rm ~~;~ if $d>4$}
\end{array} \right.
\EEQ
where $\delta(t)$ is the Dirac delta function. Eq.~(\ref{gt}) contains two kinds of terms. First, one has the expected `regular' asymptotic form, for $t\to\infty$, $g(t)\simeq f(t) M_{\rm eq}^{-4}$ and second, there appear `singular' contributions. While the response functions follow from the `regular' terms alone, the `singular' contributions are important for the correct calculation of the correlation functions, as explained in the appendix.\footnote{In \cite{Godreche00b} only the `regular' asymptotic contributions to $g(t)$ are explicitly given, while the effect of the `singular' terms in (\ref{gt}) is taken into account by non-asymptotic 
sum rules, with the same end result for the physically observable two-time correlations and responses, if $d\ne 4$.} 
In figure~\ref{Abb1}b we 
illustrate the $4D$ long-time behaviour $g(t) \sim t^{-2}$. 
We observe that {\em no} logarithmic factors occur, for $d=4$. 
Therefore, in the scaling limit $t,s\to \infty$ with $y=t/s>1$ fixed, 
we recover the simple scaling behaviour 
\BEQ \label{CRfin_ord}
C(t,s) = M_{\rm eq}^2 \left( \frac{4 y}{(y+1)^2}\right) 
\;\; , \;\;
R(t,s) =  \frac{1}{16\pi^2} \frac{1}{s^2}\frac{y}{(y-1)^2}
\EEQ
which could also have been obtained from the well-known scaling functions
for $d\ne 4$ \cite{Godreche00b} and performing the analytical continuation 
$d\to 4$. 
\\

\noindent \underline{\bf 3. $T=T_c$}. In this case, we have from (\ref{f})
\BEQ \label{gpc}
\lap{g}(p) = -\frac{1}{2 T_c} + \frac{1}{(2 T_c)^2}\left\{ \begin{array}{ll}
-a_1^{-1} p^{1-d/2}                 & \mbox{\rm ~~;~ if $2<d<4$} \\
-a_0^{-1} p^{-1} (C_E + \ln p)^{-1} & \mbox{\rm ~~;~ if $d=4$} \\
a_2^{-1} p^{-1}                    & \mbox{\rm ~~;~ if $d>4$}
\end{array} \right.
\EEQ
As shown in the appendix, for $d=4$ this leads to the long-time behaviour
$g(t)\sim 1/\ln t$. In figure~\ref{Abb1}c we illustrate that this asymptotic
form is indeed compatible with the numerical solution of (\ref{V}). 
Taking into account also the singular parts of $g(t)$, 
we find the scaling behaviour
\BEQ \label{CRfin_crit}
C(t,s) = \frac{T_c}{8\pi^2} \frac{1}{s} \sqrt{\frac{\ln t}{\ln s}\,} 
\frac{1}{y^2-1}
\;\; , \;\;
R(t,s) = \frac{1}{16\pi^2} \frac{1}{s^2} \sqrt{\frac{\ln t}{\ln s}\,}
\frac{1}{(y-1)^2} 
\EEQ
The calculations are outlined in the appendix.

Do these extra logarithmic factors imply that simple scaling is modified
in the $4D$ spherical model, at least for $T=T_c$~? Indeed, the answer is 
negative, since in the scaling limit $t,s\to \infty$ with $y=t/s>1$ fixed 
and finite, we can write $t=ys$, hence
$\sqrt{{\ln t}/{\ln s}\,}=\sqrt{{\ln (ys)}/{\ln s}\,}\simeq 
1+\demi {\ln y}/{\ln s} +\ldots$. Therefore, the scaling functions
for $d=4$ can be obtained from the known expressions for $d\ne 4$ 
\cite{Godreche00b} by analytic continuation $d\to 4$. The fact that the
system is at the upper critical dimension $d^*=4$ of its {\em equilibrium}
critical point only enters into the additive logarithmic 
corrections to scaling. In this respect, the $4D$ spherical models
shows a different behaviour from those systems 
\cite{Bray95,Rojas99,Bray00,Berthier01,Jeon03,Abriet04,Dutta05,Schehr05,Lei07,Walter07,Janssen94} 
where logarithmic corrections to non-equilibrium dynamical scaling were 
seen before. 

\subsection{Domain size}

In order to understand better the role of the upper critical dimension
$d^*=4$ for non-equilibrium dynamical scaling of the spherical model, 
we now consider the linear domain
size $L(t)$ in order to check for the presence of additional logarithmic factors
with respect to the na\"{\i}vely expected simple behaviour $L(t)\sim \sqrt{t}$,
valid for $T\leq T_c$ and $d\ne 4$. In eq.~(\ref{L2}) we had already related
$L(t)$ to the function $g(t)$ which fixes the spherical model dynamics. 
{}From this, it is already clear that for $T=0$, we simply have
$L^2(t)= 4dt$ and since the temperature $T$ is generally thought to be 
irrelevant for quenched to $T<T_c$ \cite{Bray94}, we would expect that this 
behaviour of $L(t)$ should
remain qualitatively correct for all $T<T_c(d)$. 

\begin{figure}
\centerline{
\epsfxsize=5.0in \epsffile{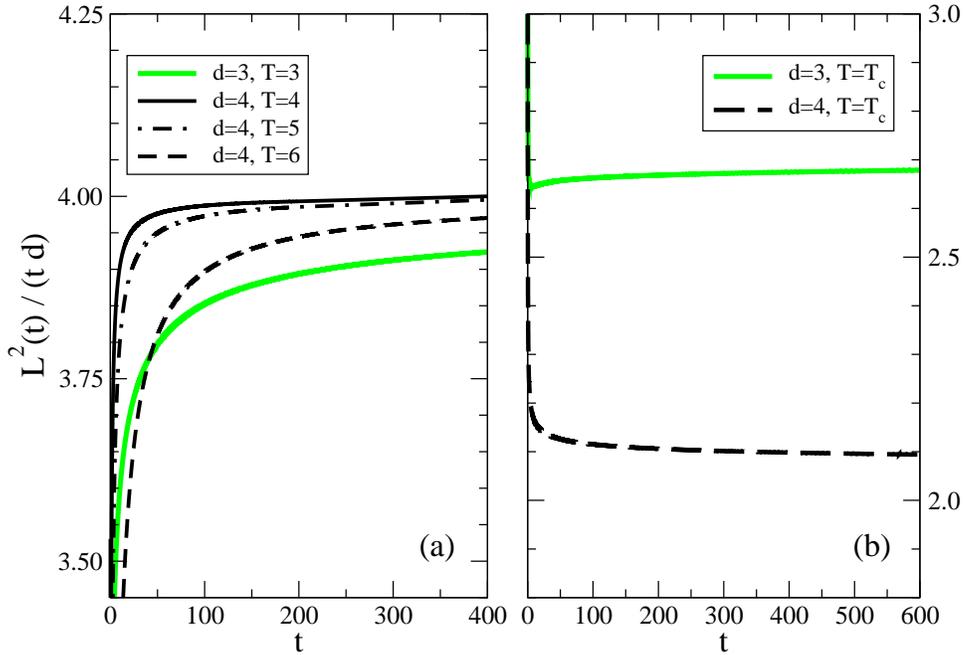}
}
\caption{Long-time behaviour of the linear domain size $L(t)$ for (a) $T<T_c$ 
and (b) $T=T_c$. Recall that $T_c(3)=3.957\ldots$ and $T_c(4)=6.454\ldots$.
\label{Abb2}
}
\end{figure}

In order to see that this expectation is indeed borne out, 
we rewrite eq.~(\ref{L2}) as the ratio of two inverse
Laplace transformations, since the two integrals can be seen as Laplacian 
convolutions
\BEQ \label{L2bis}
L^2(t) = 4d \frac{{\cal L}^{-1}(p^{-2}+2T p^{-2}\lap{g}(p))(t)}
{{\cal L}^{-1}(p^{-1}+2T p^{-1}\lap{g}(p))(t)}
\EEQ
On the other hand, for $T<T_c$, we have from (\ref{gp}) that
$\lap{g}(p)=(2T_c)^{-1} (1-T/T_c)^{-1}(1+o(p))$. Consequently, we easily find
for all $T<T_c(d)$, and for all dimensions $d>2$
\BEQ
L^2(t) = 4d t \left(1+{\rm O}(1/t)\right)
\EEQ
This is in agreement with our expectation formulated above and also with
more indirect conclusions \cite{Godreche00b} drawn from the scaling of the
single-time and two-time correlation functions. In figure~\ref{Abb2}a we 
further illustrate this by showing that the ratio $L^2(t) /(t d)\to 4$ in the
long-time limit in both $3D$ and $4D$ (in the $3D$ case, we have checked 
explicitly that convergence occurs for times $t\gtrsim 600$).

In a similar way, we now analyse the critical case $T=T_c(d)$. 
The starting point 
is again eq.~(\ref{L2bis}) where we now have to insert eq.~(\ref{gpc}). 
This leads for $t$ large to (see the appendix)
\BEQ \label{L2ter}
L^2(t) \simeq t\times \left\{ \begin{array}{ll}
8 &  \mbox{\rm ~~;~ if $2<d<4$} \\
2d\: (\ln t - 1)/(\ln t - 3/2) &  \mbox{\rm ~~;~ if $d=4$} \\
2d &  \mbox{\rm ~~;~ if $d>4$} 
\end{array} \right.
\EEQ
We see that to leading order as $t\to\infty$, we always have $L^2(t)\sim t$
and that logarithmic contribution at most enter into the additive correction
of this scaling behaviour. This shows again the difference between the $4D$ 
spherical model and the systems with time-dependent logarithmic scaling
of the form $L(t)\sim (t/\ln t)^{1/2}$ studied in the literature 
\cite{Bray95,Rojas99,Bray00,Berthier01,Jeon03,Abriet04,Dutta05,Schehr05,Lei07,Walter07}. 
In figure~\ref{Abb2}b, we illustrate this for the $3D$ and $4D$ cases. 
In the $3D$ case, a nice convergence towards the amplitude 
$L^2(t)/(t d)=8/3$ is seen, as expected from (\ref{L2ter}). 
In the $4D$ case, we extended the 
calculations up to $t=10^4$ and find $L^2(t)/(t d)\simeq 2.06$, not too
far from the approximate analytical result (\ref{L2ter}). 

We conclude that in all cases considered, we have found clear evidence that
the typical length scales as $L(t) \sim t^{1/2}$ and that there no evidence for
any logarithmic correction factors present. 
 
\section{Conclusions}

We have considered, as a case study, the ageing behaviour of the 
four-dimensional spherical model quenched to $T\leq T_c$ from a fully
disordered initial state. Since the model is at its upper critical dimension,
we had expected to find modifications of the usual scaling behaviour by
logarithmic correction factors which is indeed true when looking at the
relaxation time $\tau_{\rm eq}$ as a function of the temperature $T$, 
for $T>T_c$, and fully consistent with existing knowledge of these
corrections {\em at} equilibrium \cite{Kenna06a,Kenna06b}. Surprisingly,
when studying the time-dependent scaling, both for quenches into the
ordered phase and for quenches onto the critical point, our main
results eqs.~(\ref{CRfin_ord}, \ref{CRfin_crit}, \ref{L2ter}) show
standard simple ageing and we did not see any evidence
for logarithmic corrections appearing in the leading scaling behaviour,
although additive logarithmic corrections to scaling 
do occur for quenches to $T=T_c$, see eq.~(\ref{CRfin_crit}).
We have also found a simple power-law scaling $L(t)\sim t^{1/2}$, without
additional logarithmic factors, for the typical length scale, 
see eq.~(\ref{L2ter}). 
Surprisingly, this suggests that the fact of a system being 
at its upper critical dimension
should be considerably less relevant for its non-equilibrium dynamical behaviour
than it is for its equilibrium critical behaviour. Of course, there may
be other reasons for a system to develop logarithmic scaling, such as vortices
\cite{Bray00,Berthier01,Jeon03,Abriet04,Dutta05,Schehr05,Lei07,Walter07} 
or a roughening transition of the interfaces between ordered domains \cite{Abraham}. 

Experimentally, observations of the kind made here could be of relevance
for the non-equilibrium dynamics of systems at a {\em tri}critical point (where
$d^*=3$) and which arise for example in diluted magnets, meta-magnets or 
solutions of long polymers at a $\Theta$-point. It would be of interest to consider
a spherical meta-magnet in an external magnetic field and then study its
dynamics at the tricritical point at the upper critical dimension $d^*=3$. 
Indeed, non-equilibrium relaxation at a {\em tri}critical point was studied long
ago in the O($n$) model by Janssen and Oerding \cite{Janssen94}. At the upper
critical dimension $d=3$, starting form a small initial magnetisation, they find
to one-loop order 
the habitually expected logarithmic corrections factors, e.g. in the short-time
scaling $\langle M(t)\rangle\sim M_0 \bigl(\ln t\bigr)^{-a(n)}$ with 
$a(n)=\frac{(n+2)(n+4)}{8\pi(3n+22)}$, and similar for the correlation with
the initial state $C(t)=\langle M(t) M(0)\rangle\sim t^{-3/2} 
\bigl(\ln t\bigr)^{-a(n)}$. However, their study only considered single-time observables and
therefore cannot illustrate directly the compensation of logarithmic factors in 
{\em two}-time quantities which was found in the present work. Since $a(n)$ diverges
as $n\to\infty$, it appears also possible that the $n\to\infty$ limit might have
special properties. It remains an open problem to what extent the results on the
two-time quantities in the spherical model at $d=d^*$ reported here are generic.

\noindent
{\bf Acknowledgements:} ME and HG participate in the 
`Cursus int\'egr\'e trinational de physique SLLS (Saarbr\"ucken-Nancy-Luxembourg)' 
and are supported by the Universit\'e Franco-Allemande (UFA). 
MH thanks the Centro de F\'{\i}sica T\'eorica e 
Computaticonal of the Universidade de Lisboa for warm hospitality.

\newpage 

\appsektion{}

We outline some details of the calculation for $d=4$ whose results were quoted 
in the main text \cite{HeleMax}. For $2<d<4$ and $d>4$ all 
results can be taken over from \cite{Godreche00b}. 

First, we analyse the leading behaviour of $\lap{f}(p)$ for $p\to 0$. The first
step is to decompose the integral (we set $d=4$ from now on)
\BEQ
\lap{f}(p) = \int_0^{\infty} \!\D u\, e^{-(p+4d)u} I_0(4u)^d =: 
\lap{f}_{\rm reg}(p) + \lap{f}_{\rm sing}(p)
\EEQ
into a regular and a singular part by setting 
$\int_0^{\infty} = \int_0^{\eta} + \int_{\eta}^{\infty}$. 
The integrals are evaluated in the double limit $\eta\gg 1$, $p\ll 1$ and 
$p\eta\sim {\rm O}(1)$ kept fixed. The decomposition of the integral into
two parts is merely a heuristic device in order to arrive rapidly at the singular
terms which will be seen for $p\eta$ small enough; 
of course the end result for $\lap{f}(p)$ should be independent of
$\eta$. First, one has, 
to leading order $\lap{f}_{\rm reg}(p) \to (2T_c(4))^{-1}$. Second, we use the
identities eqs.~(3.351(4)) and (8.214(1)) from \cite{Gradshteyn80}
\BEA 
\int_{\eta}^{\infty} \!\D u\, u^{-2}\, e^{-pu} &=& p {\rm Ei}(-p\eta) 
+ \eta^{-1} e^{-p\eta} \\
{\rm Ei}(-x) &=& C_E +\ln x + \sum_{k=1}^{\infty} \frac{(-x)^k}{k! k} \;\; ; \;\;
\mbox{\rm for $x>0$}
\EEA
where ${\rm Ei}(x)$ is the exponential integral and then obtain, to leading
order in the double limit introduced above 
\BEA
\lap{f}_{\rm sing}(p) &=& \int_{\eta}^{\infty} \!\D u\, e^{-(p+16)u} I_0(4u)^4
\nonumber \\
&=& (8\pi)^{-2} \left[ p {\rm Ei}(-p\eta)  + \frac{e^{-p\eta}}{\eta} \right]
\nonumber \\
&\simeq& (8\pi)^{-2} \bigl[ p C_E + p \ln p \bigr]
\EEA
where we dropped an $\eta$-dependent term $\sim p\ln\eta$ which
should cancel anyway against a corresponding correction coming
from the regular term $\lap{f}_{\rm reg}(p)$. Then the case $d=4$ of 
(\ref{f}) follows. 

Next, we derive $\lap{g}(p)$ for $d=4$ and $T<T_c$. From the above and (\ref{glap}),
the leading terms for $p\to 0$ are
\BEQ
\lap{g}(p) \simeq \frac{1}{2T_c M_{\rm eq}^2} 
+ \frac{C_E p}{(8\pi)^2 M_{\rm eq}^4} + \frac{p \ln p}{(8\pi)^2 M_{\rm eq}^4}
\EEQ
and $g(t)$ follows by inverting the Laplace transformations, using the
identities ${\cal L}^{-1}(1)(t) = \delta(t)$ and
\BEA
{\cal L}^{-1}(p)(t) &=& \frac{\D}{\D t} \int_{c-\II\infty}^{c+\II\infty}
\!\D p\, e^{pt} \:=\:  \delta'(t) 
\nonumber \\
{\cal L}^{-1}(p\ln p)(t) &=& \frac{\D^2}{\D t^2} 
\int_{c-\II\infty}^{c+\II\infty}
\!\D p\; \frac{e^{pt} \ln p}{p} \:=\:\frac{\D^2}{\D t^2} (\ln C_E - \ln t) 
\:=\: t^{-2}
\EEA
(see eqs. (17.13.1) and (17.13.12) in \cite{Gradshteyn80}). Hence we
arrive at the expressions (\ref{gp}) and (\ref{gt}) for $d=4$. 

{}From eq. (\ref{CetR}) we get for the correlation function:
\BEQ
C(t,s) = M^4_{\rm eq}
\frac{t^{d/4}s^{d/4}}{\left(\frac{t+s}{2}\right)^{d/2}}+2TM^4_{\rm eq}(ts)^{d/4} 
\int_0^s \!\D t_1\, {f\left(\frac{t+s}{2}-t_1\right)g(t_1)}.
\EEQ
in which we estimate the value of the integral by 
developing the $f$ function at the first order:
\newpage
\BEA
&&\int_0^s \!\D t_1\, {f\left(\frac{t+s}{2}-t_1\right)g(t_1)} \nonumber \\
&\approx&f\left(\frac{t+s}{2}\right)
\int_0^s \!\D t_1\,{g(t_1)}-f'\left(\frac{t+s}{2}\right)\int_0^s 
\!\D t_1\,{t_1g(t_1)} \nonumber \\
&\approx& \left(\frac{t+s}{2}\right)^{-d/2}\frac{1}{2T_cM^2_{\rm eq}}
+d\left(\frac{t+s}{2}\right)^{-d/2}\frac{1}{t+s}
\frac{(8\pi)^{-d/2}}{M^4_{\rm eq}}\int_0^s \!\D t_1\,{t_1^{-d/2+1}}\nonumber\\
&=&  \left(\frac{t+s}{2}\right)^{-d/2}\frac{1}{2T_cM^2_{\rm eq}}+cste 
\cdot(t+s)^{-d/2-1}s^{2-d/2}
\EEA
In the second line we used the sum rule 
$\int_0^{\infty}\!\D u\, g(u)=(2T_c M_{\rm eq}^2)^{-1}$  
\cite[(eq. (2.39)]{Godreche00b}), which follows in our presentation from the first 
singular term $\sim \delta(t)$ in (\ref{gt}) and clarifies the important
contributions coming from the singular terms in $g(t)$. 
Furthermore it follows from the
explicit form of $g(t)$ that the regular long-time approximation for
$g(t)$ may be used in the second integral in the second line. 
In addition, we replace the
derivative of the function $f$ by its asymptotic value. 
In the last line, we see that the second term is negligible in the $s\to\infty$ limit. 

Therefore, using the exact expression $M_{\rm eq}^2=1-T/T_c$, we find for any dimension $d$ 
\BEA
C(t,s) &\approx& \left(\frac{4ts}{(t+s)^2}\right)^{d/4}
\left(M^4_{\rm eq}+\frac{T}{T_c}M^2_{\rm eq}\right)\nonumber\\
&=& M^2_{\rm eq}\left(\frac{4ts}{(t+s)^2}\right)^{d/4} 
\:=\: M^2_{\rm eq}\left(\frac{4y}{(y+1)^2}\right)^{d/4}.
\EEA
where the last expression gives the scaling limit $t,s\to\infty$, 
with $y=t/s$ fixed. 

The autoresponse function is found directly from (\ref{CetR}) by inserting the
long-time behaviour for $g(t)$ from (\ref{gt}). 

For the critical case $T=T_c$ at $d=4$, we have
\BEQ
\lap{g}(p) = -\frac{16\pi^2}{T_c^2}\frac{1}{C_E p+p\ln p} - \frac{1}{2T_c}
\EEQ
and we now use an approximate method based on the relation
$\ln p = \lim_{n\to 0} (p^n-1)/n$. In order to find $g(t)$, 
at least approximatively, we write for $p$ small enough
\BEQ
\frac{1}{C_E p+p\ln p} =\lim_{n\to 0}\frac{n}{n C_E-1}
\frac{p^{-1}}{(1+p^n/(n C_E-1))}
\simeq \lim_{n\to 0} \frac{n}{n C_E-1} \left( \frac{1}{p} 
- \frac{p^{n-1}}{n C_E-1} +\ldots \right)
\EEQ
Carrying out the inverse Laplace transform, we find
\BEA
{\cal L}^{-1}\left(\frac{1}{C_E p+p\ln p}\right)(t) &\simeq&
\lim_{n\to 0} \frac{n}{n C_E-1} \left( 1 - \frac{t^{-n}}{(n C_E-1)\Gamma(1-n)}
+ \ldots \right) 
\nonumber \\
&\simeq& \lim_{n\to 0} \frac{n}{n C_E-1} 
\left( 1 + \frac{e^{-n\ln t}}{(n C_E-1)\Gamma(1-n)}
+ \ldots \right)^{-1}
\nonumber \\
&=& - \frac{1}{\ln t}
\label{A:gTc}
\EEA
and hence we expect $g(t)\sim (\ln t)^{-1}$ for large times. 
The approximate derivation of (\ref{A:gTc}) relies on commuting several limits. 
In order to check whether this is justified, 
we compare in figure~\ref{Abb1}c our analytical 
result with the direct numerical solution of the Volterra equation 
and find that the main feature, namely 
the logarithmic dependence of $g(t)$ on $t$, is correctly reproduced, but
the corresponding amplitude is not. In addition, since at
criticality only ratios $g(t)/g(t')$ enter into the leading contributions
of the physical observables of interest, the corresponding amplitudes cancel and
hence will not be required.

Finally, we prove (\ref{L2ter}). For $d>4$, we consider the form
(\ref{L2bis}) together with $\lap{g}(p)= a_2^{-1}p^{-1}$. Then 
\BEQ
L^2(t) = 4d\:\frac{{\cal L}^{-1}(p^{-2}+(2T_c/a_2) p^{-3})(t)}
{{\cal L}^{-1}(p^{-1}+(2T_c/a_2) p^{-2})(t)} \stackrel{t\to\infty}{\simeq}
2d\:\frac{(2T_c/a_2)\cdot t^2}{(2T_c/a_2) \cdot t} = 2d\:t
\EEQ
since the second term in both the numerator and the denominator is the 
dominant one for $p\to 0$. 
Similarly, for $2<d<4$ we have
\BEQ
L^2(t) = 4d\frac{{\cal L}^{-1}(p^{-2}+(2T_c/a_1) p^{-1-d/2})(t)}
{{\cal L}^{-1}(p^{-1}+(2T_c/a_1) p^{-d/2})(t)}\stackrel{t\to\infty}{\simeq}
4d \frac{1/\Gamma(d/2+1)}{1/\Gamma(d/2)} t = 8t
\EEQ
Last, but not least, for $d=4$, we must find the inverse Laplace transforms of
$h_k(p) :=p^{-k} (C_E+\ln p)^{-1}$, with $k=2,3$. Applying the same procedure
we used in (\ref{A:gTc}) in order to derive $g(t)$, we obtain
\BEQ
{\cal L}^{-1}(h_2(p))(t) \simeq - \frac{t}{\ln t -1} \;\; , \;\;
{\cal L}^{-1}(h_3(p))(t) \simeq - \demi\frac{t^2}{\ln t -3/2}
\EEQ
which completes the proof. We expect that this approximation should give
the correct leading $t$-dependence. 
Since $L^2(t)$ is given by the ratio of two such 
expressions, any inaccuracy in the associated amplitudes should largely cancel.\\~\\

\noindent {\bf Note added in proof:} after this work had been completed, 
we became aware
of the paper \cite{Hase06} which studies the ageing in the spherical model with competing interactions. In particular, a detailed analysis of $g(t)$ is presented, the results of which are in agreement with our analytical findings, in particular 
$g(t)\sim 1/\ln t$ at $T=T_c$. 
\newpage



\end{document}